\documentclass[10pt,letterpaper]{article}
\usepackage{opex3}
\begin{document}
\title{
Laser-polarization-dependent photoelectron angular distributions from polar molecules}
\author{Xiaosong Zhu,$^{1}$ Qingbin Zhang,$^{1}$ Weiyi Hong,$^{1,3}$ Peixiang Lu$^{1,2}$$^{\dag}$ and Zhizhan Xu$^{1}$ }
\address{1 Wuhan National Laboratory for Optoelectronics, Huazhong University of
Science and Technology, Wuhan 430074, China\\
2 School of Science, Wuhan Institute of Technology, Wuhan 430073, China\\
3 hongweiyi@mail.hust.edu.cn}
\email{$^\dagger$Corresponding author: lupeixiang@mail.hust.edu.cn}

\begin{abstract}
Photoelectron angular distributions (PADs) of oriented polar molecules in response to different polarized lasers are systematically investigated. It is found that the PADs of polar CO molecules show three distinct styles excited by linearly, elliptically and circularly polarized lasers respectively. In the case of elliptical polarization, a deep suppression is observed along the major axis and the distribution concentrates approximately along the minor axis. Additionally, it is also found that the concentrated distributions rotate clockwise as the ellipticity increases. Our investigation presents a method to manipulate the motion and angular distribution of photoelectrons by varying the polarization of the exciting pulses, and also implies the possibility to control the processes in laser-molecule interactions in future work.
\end{abstract}
\ocis{(020.4180) Multiphoton process; (260.3230) Ionization; (270.6620) Strong-field process.}

\section{Introduction}
When atoms are irradiated by the intense lasers, strong-field processes such as above-threshold ionization \cite{Agostini,Blaga,Quan}, nonsequential double ionization \cite{Weber,Becker} and high-order harmonic generation \cite{Lewenstein,Cao,Cao2,Lan} may occur. The area of strong-field physics has developed both theoretically and experimentally in the past decades, benefiting from the advances in laser technology for intense femtosecond laser pulse generation \cite{Brabec}. Regarding the above-threshold ionization, the photoelectron angular distribution (PAD) is one of the most concerned subjects and has been studied extensively \cite{Freeman,Yang,Paulus,Zhou}. In addition to atomic ionizations, strong-field ionizations of molecules have also attracted a lot of attentions lately \cite{Tong,Pavicic,Abu-samha}, which are more complex due to the additional features such as molecular orbital symmetry, the anisotropic structure and rotational freedom of molecule. On the other side, investigations on these properties offer opportunities for many novel applications \cite{Itatani,Kamta,Lein,Meckel,Lan3}.

To embody the anisotropic structural features in the photoelectron distribution, it is necessary to fix the molecular axes adopting orientation techniques \cite{Holmegaard,Dimitrovski,Hansen,Oda} before a subsequent pulse exciting the ionization process. In previous works, linearly polarized exciting lasers are generally considered. By varying the angle between the polarization axis of laser pulse and the main axis of the orientation distribution, the ionized photoelectrons escape along the polarization axis and carry information about the objective molecule in the selected direction. For asymmetric molecules, however, information on the asymmetric structural features is not available due to the linearly polarized alternating electric fields and asymmetric PAD will not be observed in this linear polarization scheme. There are also other forms of polarizations, the circular and elliptical polarizations, which still involve abundant topics to be studied. Investigations on the PADs for both polar and non-polar molecules ionized by circularly polarized laser pulses have been performed lately \cite{Holmegaard2,Dimitrovski2,Zhu}, and very recently Abu-samha \emph{et al.} \cite{Abu-samha2} has studied the photoelectron momentum distribution of atomic oribtals by elliptically polarized laser pulses. Nevertheless, investigations on the PADs in response to elliptically polarized exciting lasers especially from polar molecules are still in request.

In this paper, the PADs of oriented polar CO molecules in response to different polarized lasers are systematically studied. The results show that the PADs for linear, elliptical and circular polarizations present three distinct styles respectively. In detail, the obtained distribution is found to be consequent on the combination of factors such as the ellipticity, the direction of the major axis and the rotation direction of the electric field. The effect of ellipticity is particularly discussed, which shows a clockwise rotation of the concentrated distributions as the ellipticity increases. Our investigation indicates a method to manipulate the motion and angular distribution of photoelectron by varying the polarization of the exciting lasers and implies the possibility to control the processes in laser-molecule interactions in future work.

\section{Theoretical model}
We carry out the calculation with the semiclassical model under the strong field approximation (SFA) \cite{Lewenstein,Keldysh,Faisal,Reiss,Lan2}. The photoelectron momentum spectrum is given by
\begin{equation}
b(\mathbf{p}) = i\int^{\infty}_{-\infty}\mathbf{F}(t')\cdot\mathbf{d}[\mathbf{p}+\mathbf{A}(t')]
exp\{-i \int^{\infty}_{t'}[(\mathbf{p}+\mathbf{A}(t''))^{2}/2+I_{p}]dt''\}dt',
\end{equation}
where $I_{p}$ is the ionization energy of the molecule, $\mathbf{p}$ is the momentum of electron, $\mathbf{F}(t)$ is the electric field of the laser pulse and $\mathbf{A}(t)$ is the vector potential.

In Eq. (1), we integrate through the whole time when the laser pulse interacts with atoms or molecules, so the final momentum distribution after the interaction is obtained. $\mathbf{d}(\mathbf{p})$ is the dipole moment, which is calculated by \cite{Etches,Faria}
\begin{equation}
\mathbf{d}(\mathbf{p})=i\nabla_{\mathbf{p}}\widetilde{\Psi}(\mathbf{p}).
\end{equation}
$\widetilde{\Psi}(\mathbf{p})$ is the Fourier transform of the highest occupied molecular orbital (HOMO) $\Psi(\mathbf{r})$ in coordinate space, which is obtained with the Gaussian 03 $\emph{ab initio}$ code \cite{Gaussian}. Eq. (2) is derived from the formula
\begin{equation}
\mathbf{d}(\mathbf{p})=\frac{1}{(2\pi)^{3/2}}\int d^3r\mathbf{r}exp[-i\mathbf{p}\cdot\mathbf{r}]\Psi(\mathbf{r}),
\end{equation}
where plane waves are used as the continuum state. The plane wave approximation is appropriate to some extent that only the overall angular distribution of the momentum spectrum is concerned in this work. For investigations of fine structures, Coulomb continuum wave functions are needed \cite{Ciappina}.

\section{Result and discussion}
In this paper, we focus on the polar CO molecules, which are oriented parallel to the x axis in the laboratory frame with C to the -x direction and O to the +x direction. The ionization energy of CO is 14.01 eV. Through our work, we apply 532 nm exciting pulses which are the second harmonic of 1064 nm pulses from solid infrared laser. A relative high intensity of 0.4 PW/cm$^2$ is considered, because for nonlinear polarizations the amplitude of electric field is lower than that in linear polarization and sufficiently high electric field is needed to satisfy the requirement of SFA theory. The duration of the laser pulse is 10 optical cycles, where a trapezoidal profile with 3-cycle ramps is adopted. In the plateau region of a multi-cycle pulse, the electric field is
\begin{eqnarray}
&&F_x(t)=F_{0x}cos(-\omega t+\varphi_0),\\
&&F_y(t)=F_{0y}cos(-\omega t+\varphi_0+\delta),
\end{eqnarray}
where the amplitudes of the two components $F_{0x}=F_{0y}$. The polarization form of the laser pulse is determined by $\delta$, which is the phase difference between the two components. For example, the pulse is linearly polarized when $\delta=0$, and it goes left or right circularly polarized when $\delta$ turns to $\pi/2$ and $-\pi/2$ respectively.

As a special case of elliptical polarization, we first investigate the PAD of CO in response to circularly polarized laser pulses, where both left circular polarization (LCP) and right circular polarization (RCP) are under consideration. The obtained photoelectron momentum spectra are presented in Fig 1. In the top row, panels (a) and (b) show the spectra for LCP and RCP respectively without including the Stark shift of CO \cite{Dimitrovski2}, while in (c) and (d) the effect of Stark shift is considered. In the middle of the figure, the geometry and layout of the CO molecule is shown in the inset.

\begin{figure}[htb]
\centerline{
\includegraphics[width=12cm]{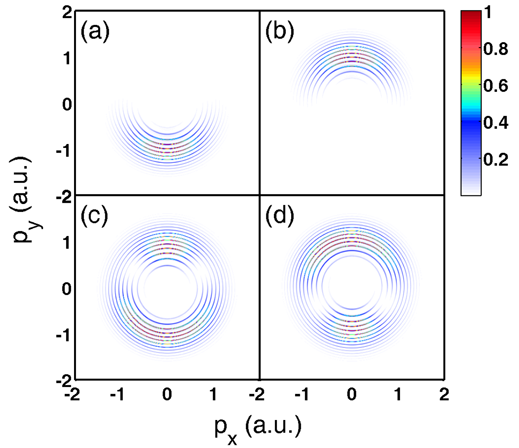}}
 \caption{Photoelectron momentum spectra for CO in the cases of (a) LCP pulse without Stark shift, (b) RCP pulse without Stark shift, (c) LCP pulse with Stark shift, (d) RCP pulse with Stark shift respectively .}
\end{figure}

In the spectra, we can see typical stripes with spacings between adjacent ones equivalent to absorbed photon energy, which is corresponding to the well known ATI (above-threshold ionization) peaks. In panels (a) and (b), crescent stripes are observed instead of rings because of the asymmetric structure of CO and the consequent asymmetric ionization rate. As illustrated in Fig 2, since the electron density concentrates around the C nucleus in the highest occupied molecular orbital (HOMO), ionization rate when the electric field points to the +x direction (solid green arrow) is much higher than that when electric field points to the -x direction (dashed blue arrow). After ionized at time $t_0$, the electrons are still driven by the remaining part of laser pulse, and the final momentum is given according to the simple-man model \cite{Holmegaard2,Dimitrovski2,Corkum}
\begin{equation}
\mathbf{p}=-|e|\int_{t_0}^{\infty}\mathbf{F}(t)dt=-|e|\mathbf{A}(t_0),
\end{equation}
i.e. the final escape direction of the electron is right opposite to the transient direction of the vector potential $\mathbf{A}(t)$ at the ionization moment $t_0$. According to the definition of vector potential, we can find the relationship
\begin{eqnarray}
\mathbf{A}(t)=\int_{t}^{\infty}-\mathbf{F}(t')dt'=\frac{F_{0x}}{\omega}sin(-\omega t+\varphi_0)\mathbf{\hat{x}}+\frac{F_{0y}}{\omega}sin(-\omega t+\varphi_0+\delta)\mathbf{\hat{y}},\\
\frac{d}{dt}\mathbf{F}(t)=\omega F_{0x}sin(-\omega t+\varphi_0)\mathbf{\hat{x}}+\omega F_{0y}sin(-\omega t+\varphi_0+\delta)\mathbf{\hat{y}},
\end{eqnarray}
so the transient direction of vector potential is the same as $\frac{d}{dt}\mathbf{F}(t)$. As a result, in Fig 2 where an LCP pulse is taken as an example, the directions of vector potentials are 90 degree rotated to the left with respect to the electric fields. Thus the directions of final momenta of the emitted electrons are obtained opposite to the vector potentials. Fig 2 has well illustrated the observed photoelectron distribution observed in Fig 1(a): electrons are dominantly ionized at the instant when electric field points to the +x direction, the ionized electrons will then be driven by the remaining part of the pulse and finally escape along the -y direction (see the green arrows and notations). This explains the crescent stripes concentrated in the bottom half of Fig 1(a). On the contrary, the ionization rate is much lower when the electric field points to the -x direction and few electrons will distribute along the +y direction consequently (see the blue arrows and notations). Regarding Fig 1(b), the vector potentials and final momenta reverse their directions, so the observed crescent stripes distribute in the top half of the spectrum.

\begin{figure}[htb]
\centerline{
\includegraphics[width=6cm]{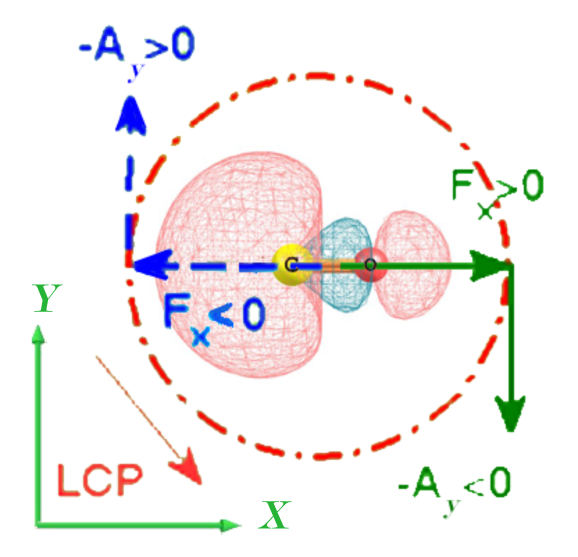}}
\caption{ Illustration for the photoelectron angular distribution of CO in response to circularly polarized laser pulses.}
\end{figure}

In Fig 1(c) and (d) we introduce the effect of Stark shift, which will lead a time-dependent ionization energy in respond to the electric field owing to the permanent dipole of polar molecules. Considering the linear Stark shift of HOMO, the time-dependent ionization energy is given by \cite{Abu-samha,Etches}
\begin{equation}
I_p(t)=I_{p0}+\mathbf{\mu}_h\cdot\mathbf{F}(t),
\end{equation}
where $I_{p0}$ is the field free ionization energy, $\mathbf{\mu}_h$ is the permanent dipole of HOMO of CO and $\mathbf{F}(t)$ is the external field. Regarding the validity of this modification, Dimitrovski \emph{et al.} \cite{Dimitrovski2} have stressed that the applicability of this Stark-shifted MO-SFA is restricted to the tunneling limit. In our case, the Keldysh parameter $\gamma=\omega\sqrt{2I_p}/F$ varies from 0.815 to 1.152 with the polarization changing from linear to circular, namely the ionizations are in or very close to the tunneling regime and the tunneling is dominating \cite{Eckle}. The permanent dipole of HOMO is calculated to be 1.72 a.u. (4.37 D) by \cite{Abu-samha}
\begin{equation}
\mathbf{\mu}_h=-\int d\mathbf{r}\mathbf{r}\rho^H(\mathbf{r}),
\end{equation}
where $\rho^H(\mathbf{r})$ is the electron density of the HOMO calculated by $\rho^H(\mathbf{r})=\int d\mathbf{r}\Psi^*(\mathbf{r})\Psi(\mathbf{r})$.

It has been pointed out that the ionization of CO will be decreased when the electric field is parallel to the permanent dipole while will be increased in the antiparallel geometry due to the Stark shift, which will offset the effect of asymmetric ionization rate discussed above \cite{Etches}. Similarly, comparing Fig 1(c)(d) with (a)(b), there are still photoelectron distributions on the opposite side of those crescent stripes, which results from the fact that the Stark shift has enhanced the ionization when the electric field points to the -x direction leading additional photoelectron distribution along +p$_y$ axis (see dashed blue arrow in Fig 2). As the effect of Stark shift is not negligible for CO who has considerable permanent dipole, it will be contained in all of the simulations in the following of this paper. The present analysis method associated with the simple-man model and the Stark shift has been developed by L. B. Madsen and coworkers for analyzing the photoelectron distributions from atoms to complex molecules in their theoretical works \cite{Dimitrovski2,Abu-samha2}, and has also successfully been applied to explain the their observations in the experiments on PAD detections from oriented molecules \cite{Dimitrovski,Hansen,Holmegaard2}.

Next, we will systematically investigate the PAD of CO in response to various forms of polarized laser pulses. Eight typical values of phase differences $\delta$ in the interval between -$\pi$ and $\pi$ are chosen, corresponding to different polarizations from linear, elliptical to circular ones and different rotation directions. The obtained photoelectron momentum spectra are presented in Fig 3, with the x-y views of applied electric fields sketched in the insets of respective panels.

\begin{figure}[htb]
\centerline{
\includegraphics[width=14cm]{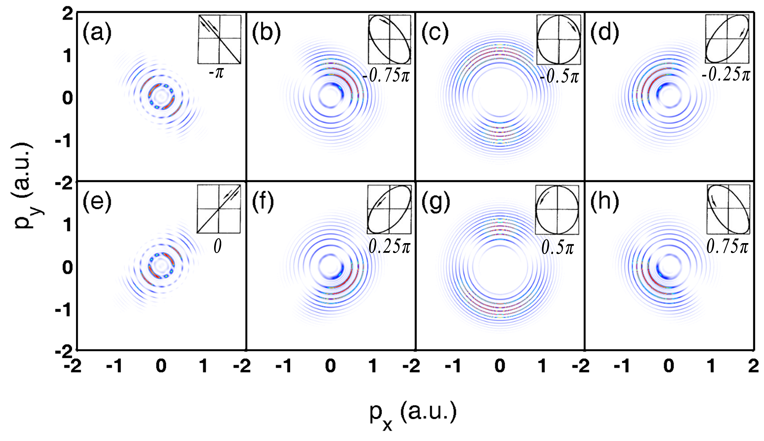}}
\caption{(Color online) Photoelectron momentum spectra of CO in response to 8 typical forms of polarized laser pulses with $\delta$ varying in the interval from $-\pi$ to $\pi$. The x-y views of applied electric fields are sketched in the insets of respective panels.}
\end{figure}

In both panels (a) and (e) where the pulse is linearly polarized, the distributions concentrate along the polarization axes symmetrically. No asymmetric distribution is observed in these spectra, although the objective molecule is asymmetric. That is because we are using multi-cycle pulses so that the asymmetrically ionized electrons are averaged by the linear alternating electric filed. In detail, considering the linearly polarized laser pulse with long duration, the electric field in the plateau region is $F(t)=F_{0} cos(-\omega t+\varphi_0)\mathbf{\hat{n}}$, where $\mathbf{n}$ indicates the polarization direction. The time-dependent $F(t)$ and the colinear vector potential $A(t)=\frac{F_{0}}{\omega}sin(-\omega t+\varphi_0)\mathbf{\hat{n}}$ are both plotted in Fig 4. The ionization rate peaks at $t_1$ when the electric field peaks in the $+\mathbf{n}$ direction, at which time the vector potential $A(t_1)$ equals 0. This means electrons ionized at $t_1$ will finally gain zero momentum according to Eq (6). At moments $t_2$ and $t_3$ a little earlier and later than $t_1$, the ionization rates are the same due to equal electric fields, however the vector potentials $A(t_1)$ and $A(t_2)$ are just in opposite directions. That is to say, in each half optical cycle the ionized electrons are split equally into two parts, escaping to opposite directions $+\mathbf{n}$ and $-\mathbf{n}$ respectively. That is why symmetric distributions are finally obtained in spite of using asymmetric objective molecule.

\begin{figure}[htb]
\centerline{
\includegraphics[width=13cm]{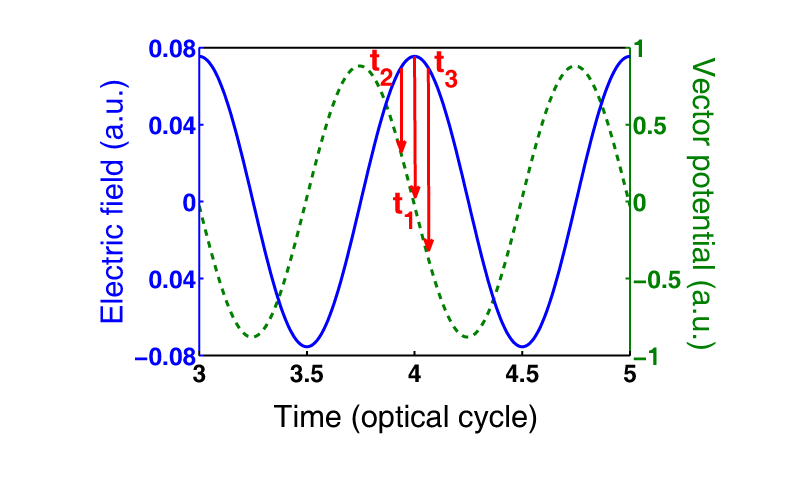}}
\caption{(Color online) The time-dependent electric field (solid blue curve) and vector potential (dashed green curve) of a linear polarized pulse in the plateau region. $t_1$ to $t_3$ are three ionization times in the same half optical cycle close to the peak of electric field, where $t_1$ corresponds to zero vector potential and $t_2$, $t_3$ correspond to two opposite vector potentials.}
\end{figure}

In the rest of the six panels, it is found that the asymmetric features are distinguished by nonlinearly polarized pulses. The difference between the linear and nonlinear polarizations is that the electric field rotates its direction gradually in nonlinear polarization instead of reversing the direction immediately at the turning point, so the electrons ionized in the same half optical cycle will gain final momenta with close directions instead of splitting into two equal parts with opposite directions.

We have already discussed the response to circularly polarized pulses (panels (c) and (g)), furthermore for the elliptical polarization the case is much more complex with respect to different rotation directions and directions of major axes (panels (b)(d)(f) and (h)). As a result of the combination of these two factors, the main distributions orient to 4 different directions. The observation in these four panels can be resolved into two main features: the deep suppression along the major axis of the ellipse and asymmetry of pairs of unequal crescent stripes.

To study the process in detail, we focus on the instance $\delta=0.25\pi$. The time-dependent and angle-dependent MO-ADK rates are calculated \cite{Tong}, which are shown in Fig 5(a) and (b). The ionization rate peaks when the electric field is approximately parallel to the major axis ($\sim 43^\circ$). There is also a secondary peak in the opposite direction and ionizations perpendicular to the major axis is nearly completely suppressed. Due to the fact that nearly no electron ionize at perpendicular directions (the dashed blue arrow of $F(t_2)$ in Fig 5(c)) compared with those at parallel directions (the solid blue arrow of $F(t_1)$ in Fig5(c)), few electrons will finally escape along the major axis, which explains the deep suppressions in all of the four spectra. Figure 5(d) illustrates the other observation that the distribution stripes at one side is stronger than the other side. Similarly, since the ionization rate maximizes when the direction of electric field is at around 43$^\circ$, photoelectrons ionized at these moments will gain final momenta 90$^\circ$ rotated on the right and lead the major distribution in the fourth quadrant (see the solid green arrows in Fig 5(d)). On the other hand, ionizations at around -137$^\circ$ will result in the minor distribution in the second quadrant (see dashed blue arrows in Fig 5(d)), which is a consequence of the Stark shift. Although it has been discussed that the Stark shift will offset the asymmetric distribution, it is shown that this effect is modest.

\begin{figure}[htb]
\centerline{
\includegraphics[width=11cm]{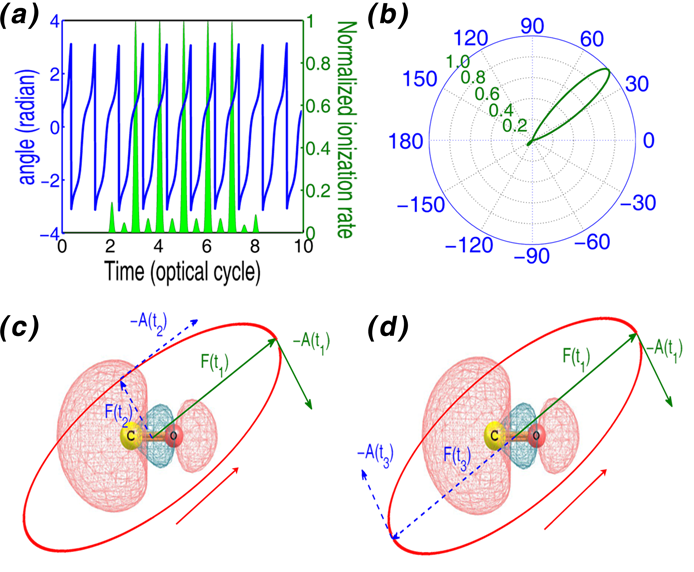}}
\caption{(Color online) (a) Time-dependent and (b) angle-dependent ionization rates calculated by MO-ADK theory. (c) Illustration for the suppression along the major axis in the photoelectron momentum spectrum with $\delta=0.25\pi$. (d) Illustration for the asymmetry of pairs of unequal crescent stripes.}
\end{figure}

To make it clearer how the photoelectron momentum spectrum evolves with the variation of polarization from linear to circular, we choose another group of eight phase differences $\delta$ in the interval between 0 and 0.5$\pi$. In Fig 6, we can see the evolution step by step from panels (a) to (g). In panel (a) with $\delta=0.1\pi$, the obtained photoelectron momentum spectrum is quite similar to that in response to linearly polarized pulse. The distribution concentrates along the major axis and only very week asymmetry can be seen in the spectrum. It means that the elliptically polarized laser pulse behaves similarly to the linearly polarized pulse when the ellipticity is low. Electrons are ionized when the electric field is parallel to the major axis and then driven basically forward and backward along the axis by the pulse. As the ellipticity grows bigger, typical spectra in response to elliptically polarized pulses are observed in panels (b) and (c). Distribution along the major axis is now suppressed instead of preferred, which has been explained previously. With the ellipticity continuing to increase, suppression along the major axis disappears again and is replaced by the suppression parallel to the x axis. In addition to that, it is observed that the concentrated angular distribution rotates clockwise gradually during the evolution from elliptical polarization to circular polarization. Finally the result tends to be the same as that in the case of circular polarization.

\begin{figure}[htb]
\centerline{
\includegraphics[width=14cm]{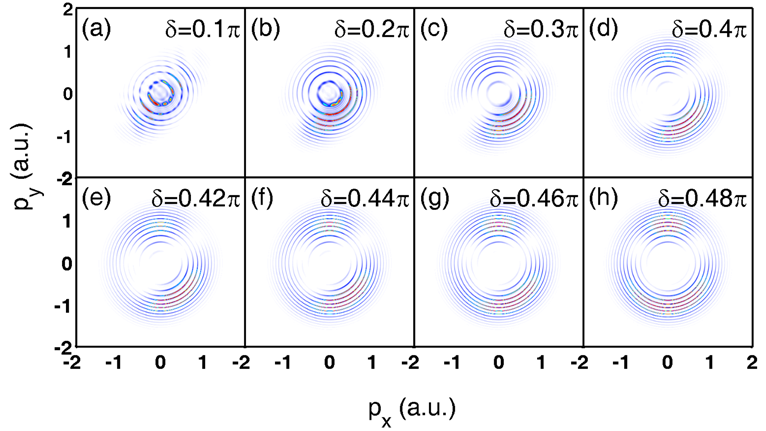}}
\caption{(Color online) Photoelectron momentum spectra of CO in response to another group of 8 polarized laser pulses with $\delta$ varying in the interval from 0 to $0.5\pi$.}
\end{figure}

So far, we have seen the full process of the photoelectron distribution changing among the three distinct styles. To understand the underlying reason, normalized angle-dependent MO-ADK rates for the eight $\delta$ values are calculated as shown in Fig 7. The ionization rate is considerable only in a very narrow range of angles at $\delta=0.1\pi$ and $\delta=0.2\pi$, quite similar to linear polarization. Then the contours of the curves become fatter and fatter as $\delta$ increases, which is because the contour of the x-y view of the electric field is going fatter and more isotropic with $\delta$ changing from 0 to $0.5\pi$. This explains the observation that the bigger the ellipticity is the smaller the contrast is in the spectrum. Besides, it is found in Fig 7 that the peak of angle-dependent ionization rotates clockwise with the increase of ellipticity. When the ellipticity is small, the electric filed is much stronger along the major axis than the minor axis, as a result the angular distribution of ionization is predominated by the electric field, an evidence of which is the ionization rate peaks at about 43$^\circ$ in Fig 5(b) and the PAD peaks around -45$^\circ$ in Fig 3(f). At big ellipticity when the electric field is tending isotropic, as the ionization rate is the combined result of electric field and the structural features of the molecule itself, the latter factor come to be important and draws the peak rotating to the +x direction. Based on the above discussion for evolution of angle-dependent ionization rate and the relationship $\mathbf{p}=-|e|\mathbf{A}(t_0)\sim\frac{d}{dt}\mathbf{F}(t_0)$, the observed clockwise rotation of concentrated distribution and the suppressions in the spectra can be explained similarly to Figs 2 and 5, that the rotating ionization peak leads the rotation of the major photoelectron distribution which is on the right side with respect to the direction of the electric field.

There is still one thing to be noted, the Stark shift is also an influential factor affecting the angular distribution of ionization, which is not included in the MO-ADK calculation. According to the obtained photoelectron spectra from semiclassical simulation, for example in Fig 3(f) where the secondary PAD peak is at about 135$^\circ$, it is shown that the ionization direction is also predominated by the electric field at low ellipticity and the ionization secondarily peaks at around -135$^\circ$ (refer to Fig 5(d)). Similarly, the Stark shift will come to be important for the angular distribution of ionization when the ellipticity is big and will draw the secondary peak rotating to the -x direction. From the panels in Fig 6, it is found that the rotation of the two concentrated distributions are not synchronized, which is right because the rotation of the two distributions are due to the two different reasons.

\begin{figure}[htb]
\centerline{
\includegraphics[width=11cm]{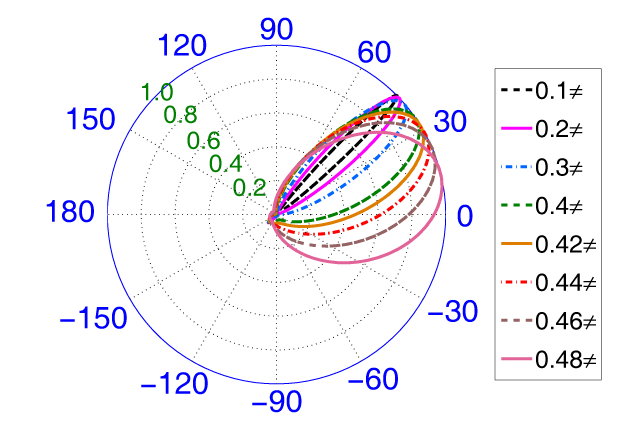}}
\caption{(Color online) Polar plot of normalized angle-dependent MO-ADK ionization rates for the eight phase differences $\delta$ in Fig 6. }
\end{figure}

We have systematically investigated the PADs of CO in response to different polarized pulses and have shown that the polarizations of the exciting pulses play very important roles on the distributions. We have also in detail discussed the dependence of the distribution to the factors that affect. Taking advantage of this sensitivity, one will be able to manipulate the photoelectron distributions by choosing proper forms of exciting laser pulses. The first thought method is to control the phase difference $\delta$, which is shown to be very effective. Besides, it is also worth considering to adjust the ratio $F_{0x}/F_{0y}$ between the two components and to rotate their directions with respect to the molecule frame. Different molecules may have different PAD responses because of different molecular features, but the mechanism remains the same. Being able to manipulate the motion and final distribution of photoelectrons might promise the possibility to control the processes in laser-molecule interactions in future work.

To control and to detect are usually two accompanying sides for one thing, thus we can think about molecule detecting via photoionization by multi-cycle polarized laser pulses at the end of this paper. Previous discussion has shown that a linearly polarized multi-cycle pulse is not able to detect the asymmetric feature of a polar molecule while nonlinearly polarized pulses have the potential. In the circular polarization condition, the electric field is constant in any directions as long as the pulse is long enough, so the ionization rate only depends on the features of molecule itself regardless of the direction of the electric field. That is the reason why the circularly polarized pulses can be used for molecular orbital imaging in one-shot scheme \cite{Zhu}. Whereas, in the case of elliptical polarization with modest ellipticity, ionizations are dominant at particular angles approximately along the major axis. That is to say, only molecular features in these directions can have effective influence on the ionization and can be encoded in the photoelectron momentum spectra. This inspire us that one may be able to acquire the asymmetric structural information of objective molecule along a selected direction with elliptically polarized multi-cycle pulse by rotating the major axis.

\section{Conclusion}
PADs of oriented polar CO molecules in response to different polarized pulses have been systematically investigated in this paper. Eight typical forms of polarizations with phase differences $\delta$ from $-\pi$ to $\pi$ are employed, the results show that the PADs excited by linearly, elliptically and circularly polarized pulses have three distinct styles respectively. For linear polarization, the distribution concentrates along the polarization direction, which is still symmetric although asymmetric objective molecules are applied on condition that the pulse duration is long enough. For circular polarization, the distribution is perpendicular to the internuclear axis and the direction of the distribution is also dependent on the rotation direction of the electric field. While in the case of elliptical polarization, the PAD is determined by the combination of the ellipticity, the direction of major axis and the rotation direction of the electric field. For typical elliptically polarized pulse with intermediate ellipticity, the distribution concentrates approximately perpendicularly to the major axis and is deeply suppressed along the this axis. Furthermore, the effect of ellipticity is particularly concerned. Another group of 8 polarizations with $\delta$ varying from 0 to 0.5$\pi$ are employed, it is found that the concentrated distributions rotate clockwise with the increase of ellipticity in the evolution of polarization from elliptical to circular. Our investigation indicates a method to manipulate the motion and final distribution of photoelectrons in the photoionization by varying the polarization of the exciting lasers, and also implies the possibility to control the processes in laser-molecule interactions in future work. In the end, we have briefly discussed the possibility to detect the asymmetric spatial features in the selected direction with proper polarized laser pulses.

\section*{Acknowledgment}
This work was supported by the National Natural Science Foundation
of China under Grants No. 60925021, 10904045, 10734080 and the National
Basic Research Program of China under Grant No. 2011CB808103. This
work was partially supported by the State Key Laboratory of
Precision Spectroscopy of East China Normal University.
\end{document}